\documentclass[conference]{IEEEtran}
\IEEEoverridecommandlockouts
\usepackage{cite}
\usepackage{amsmath,amssymb,amsfonts}
\usepackage{algorithmic}
\usepackage{graphicx}
\usepackage{textcomp}
\usepackage{xcolor}
\def\BibTeX{{\rm B\kern-.05em{\sc i\kern-.025em b}\kern-.08em
    T\kern-.1667em\lower.7ex\hbox{E}\kern-.125emX}}
\begin{document}

\title{How Developers Discuss Generative AI: A Longitudinal Study of the Visual Studio Code Community}
%\title{The Evolution of Developer Discussions on Generative AI in Open Source Projects: A Longitudinal Topic-Modeling Study of Visual Studio Code}

\author{\IEEEauthorblockN{1\textsuperscript{st} Panida Rumriankit}
\IEEEauthorblockA{\textit{Faculty of Engineering} \\
\textit{Kasetsart University}\\
Bangkok, Thailand \\
panida.rum@ku.th}
\and
\IEEEauthorblockN{2\textsuperscript{nd} Akito Monden}
\IEEEauthorblockA{\textit{
Faculty of Environmental, Life, Natural Science and Technology} \\
\textit{Okayama University}\\
Okayama, Japan \\
monden@okayama-u.ac.jp}
\and
\IEEEauthorblockN{3\textsuperscript{rd} Hiroki Inayoshi}
\IEEEauthorblockA{\textit{
Faculty of Environmental, Life, Natural Science and Technology} \\
\textit{Okayama University}\\
Okayama, Japan \\
inayoshi@okayama-u.ac.jp}
\and
\IEEEauthorblockN{4\textsuperscript{th} Pattara Leelaprute}
\IEEEauthorblockA{\textit{Faculty of Engineering} \\
\textit{Kasetsart University}\\
Bangkok, Thailand \\
pattara.l@ku.ac.th}
\and
\IEEEauthorblockN{5\textsuperscript{th} Bundit Manaskasemsak}
\IEEEauthorblockA{\textit{Faculty of Engineering} \\
\textit{Kasetsart University}\\
Bangkok, Thailand \\
bundit.m@ku.th}
\and
\IEEEauthorblockN{6\textsuperscript{th} Kundjanasith Thonglek}
\IEEEauthorblockA{\textit{Faculty of Engineering} \\
\textit{Kasetsart University}\\
Bangkok, Thailand \\
kundjanasith.th@ku.th}
\and
\IEEEauthorblockN{7\textsuperscript{th} Arnon Rungsawang}
\IEEEauthorblockA{\textit{Faculty of Digital Technology} \\
\textit{Chitralada Technology Institute}\\
Bangkok, Thailand \\
arnon.run@cdti.ac.th}
}

\maketitle

\begin{abstract}
Generative AI tools such as GitHub Copilot, ChatGPT, and coding agents have rapidly become part of everyday software development, yet little is known about how mainstream open source communities discuss them in practice. This paper presents a longitudinal analysis of generative-AI-related discussions in the Visual Studio Code (VS Code) GitHub repository, using 43,806 candidate issues created between January 2021 and June 2026. To improve corpus quality, we combined keyword retrieval with semantic relevance filtering, yielding a filtered corpus of 25,227 AI-related issues. We applied BERTopic to the retrieved corpus to identify discussion topics, using the filtered corpus for theme validation and a robustness re-clustering, and analyzed their evolution over time using monthly prevalence and Mann--Kendall trend tests.
The results show that developer discussions are dominated by practical concerns regarding the operation of AI-assisted development environments, including agent management, configuration, reliability, authentication, and billing, whereas risks frequently emphasized in survey-based studies, such as hallucination and licensing, rarely surface in this venue.
%This suggests that the issue tracker of a project with a first-party AI integration functions primarily as an operational support channel.
This suggests that discussions of generative AI in the VS Code issue tracker primarily focus on operational aspects of AI-assisted software development.
Furthermore, discussions evolved from AI-assisted code completion toward conversational and agent-based development, reflecting the increasing integration of generative AI into software development workflows. These findings suggest that GitHub Issues provide a practical, workflow-oriented perspective on generative AI that complements survey-based studies of developer perceptions.
\end{abstract}


\section{Introduction}
The rapid rise of generative artificial intelligence (AI) tools---including ChatGPT, GitHub Copilot, Google Gemini, and agentic coding systems---has fundamentally altered how software developers work. These tools assist with code generation, documentation, testing, bug fixing, and code review, reshaping the daily practices of millions of developers.

Open source software (OSS) communities provide a particularly valuable opportunity to study this phenomenon.
OSS repositories preserve large-scale, naturally occurring discussions among developers, allowing researchers to observe how perceptions and concerns evolve throughout the adoption of new technologies.

Prior empirical work has quantified the productivity impact of AI assistants \cite{b1}, \cite{b2}, catalogued their risks \cite{b3}, \cite{b4}, \cite{b21}, \cite{b22}, and mined AI-specialist repositories for discussion topics \cite{b8}, \cite{b9}, \cite{b10}.
However, relatively little is known about how developers in a mainstream software project discuss generative AI during its rapid evolution
---from Copilot as an optional completion extension, through conversational chat, to autonomous agent modes.

To address this gap, we conduct a longitudinal empirical study of AI-related discussions in the Visual Studio Code (VS Code) repository, one of the largest and most active software development communities on GitHub. By analyzing GitHub Issues from January 2021 through June 2026, we investigate not only what developers discuss regarding generative AI, but also how these discussions have changed throughout the evolution of AI-assisted software development.

Specifically, we investigate the following research questions.

\begin{itemize}
\item \textbf{RQ1:} What topics related to generative AI emerge in discussions within the VS Code community?
\item \textbf{RQ2:} What concerns and expectations do these discussions reveal regarding the use of generative AI?
\item \textbf{RQ3:} How have these discussions evolved over time from 2021 to 2026?
\end{itemize}

Our contributions are threefold.

\begin{itemize}
\item We provide, to the best of our knowledge, the first longitudinal characterization of generative-AI-related discussions in a mainstream OSS community, covering the
%complete
transition from AI-assisted code completion to conversational assistants and agent-based software development.

\item We reveal that developer discussions in this venue are dominated by operational concerns---including AI infrastructure, configuration, authentication, quotas, billing, and tool reliability---whereas conceptual risks emphasized in previous research, such as hallucinations, licensing, and code quality, appear infrequently in issue-tracker discourse.

\item We demonstrate that these findings are supported by a robust empirical analysis pipeline combining semantic relevance filtering, topic modeling, manual validation, and temporal trend analysis, enabling reliable identification of long-term changes in developer discussions.
\end{itemize}

\section{Related Work}

\subsection{Developer Use and Perception of Generative AI}
Early studies focused on productivity: Peng et al. \cite{b1} found developers using Copilot completed an implementation task 55.8\% faster, and Song et al. \cite{b2} measured a 5.9\% increase in project-level code contributions alongside an 8\% increase in coordination time in OSS projects. On the risk side, Liu et al. \cite{b21} established a taxonomy of hallucinations in LLM-generated code; Fu et al. \cite{b4} found security weaknesses in roughly 30\% of AI-generated snippets across 43 CWE categories; GitClear \cite{b3} reported a fourfold growth in code clones since the rise of AI assistance; and Xu et al. \cite{b22} showed top LLMs reproduce code strikingly similar to existing open source implementations, often without license information. Tufano et al. \cite{b5} studied self-admitted AI use in OSS commits, issues, and pull requests, deriving a taxonomy of 64 usage tasks; their reliance on explicit disclosure, however, likely underestimates the breadth of AI-related discussion, motivating the keyword-based mining used here.

\subsection{Mining GitHub Issues}
GitHub Issues are a well-established data source in empirical software engineering. Yang et al. \cite{b8} analyzed 24,953 issues in AI research repositories, finding runtime errors and unclear instructions to be the dominant categories and observing that only 7.81\% of repositories label their issues---necessitating text-based analysis over metadata filtering. Lin et al. \cite{b9} applied LDA to 23,609 issues from ChatGPT-related projects, discovering ten topics whose prominence shifts over project lifecycles; Asgari et al. \cite{b10} mined AI-agent frameworks across Stack Overflow and GitHub. All three target AI-specialist communities, leaving open how mainstream developers discuss generative AI---the population this study addresses. Han et al. \cite{b20} found that GitHub Issues capture operational, workflow-oriented concerns, which our results strongly corroborate.

\subsection{Topic Modeling in Software Engineering}
LDA \cite{b14} remains the dominant topic-modeling technique in software engineering \cite{b17}, with Barua et al. \cite{b15} establishing the canonical workflow of topic extraction plus temporal trend analysis that we extend. Treude and Wagner \cite{b16} showed that LDA requires corpus-specific configuration, a caution we follow. BERTopic \cite{b18} combines transformer sentence embeddings, UMAP, HDBSCAN, and class-based TF-IDF, capturing contextual similarity that bag-of-words models miss---particularly valuable for short, jargon-dense issue text---at the cost of sensitivity to embedding and clustering hyperparameters. Silva et al. \cite{b17} recommend coherence scores plus manual inspection with Cohen's Kappa as the standard evaluation suite, which we adopt.

\section{Methodology}

\subsection{Data Collection}\label{sec:collect}
We collected issues from microsoft/vscode using the GitHub Search API with twelve case-insensitive keywords: ChatGPT, GPT, Copilot, Claude, Cursor, Gemini, LLM, generative AI, gen AI, AI-generated, AI-assisted, and AI assistant. Because the Search API caps any query at 1,000 results and limits boolean operators, we issued one query per keyword per monthly window from January 2021 to June 2026, recursively splitting windows that exceeded the cap and deduplicating by issue number. Collection was performed on 9 June 2026, so the final month is partial (issues created 1--9 June 2026). The term ``GPT'' was excluded from server-side queries due to substring false positives (e.g., ``Egypt'') and applied only as a client-side post-filter. For each matching issue we retrieved the title, body, labels, state, timestamps, and all comments, retaining issues where at least one keyword appears in the title, body, or any comment. Pull requests were excluded. This high-recall keyword stage yielded 43,806 issues.

Each document consists of the issue title, body, and comments. Standard text preprocessing was applied, including removal of code fragments, HTML tags, URLs, and other non-textual artifacts, followed by lemmatization and stop-word removal. Documents containing fewer than five tokens were discarded, reducing the corpus from 43,806 to 43,761 documents. We refer to these 43,761 preprocessed documents as the keyword corpus.

Since keyword matching alone retrieved many editor-related issues that only incidentally mentioned AI-related terms, we constructed a semantic relevance classifier using MiniLM sentence embeddings and logistic regression trained on 300 manually labeled issues. The classifier achieved a precision of 0.79 and a recall of 0.84 under five-fold cross-validation. Applying this classifier to the keyword corpus produced a filtered corpus of 25,227 AI-related issues (57.6\%). The primary topic model and trend analysis were fit on the keyword corpus; the filtered corpus is used for the theme distribution and for the robustness re-clustering reported in Section~\ref{sec:rq3}.

\subsection{Topic Modeling}
Our primary method is BERTopic \cite{b18} with all-MiniLM-L6-v2 sentence embeddings, UMAP dimensionality reduction, HDBSCAN clustering (minimum topic size 50), and a unigram--bigram class-based TF-IDF representation; outlier documents were reassigned to their nearest topic by embedding similarity.

For comparison, we also trained an optimized LDA model following Treude and Wagner \cite{b16}. The number of topics ($k=5$--30) was selected using the $C_v$ coherence score.

\subsection{Validation and Temporal Analysis}

Topic quality was evaluated using three complementary approaches. First, quantitative quality measures ($C_v$, $C_{\mathrm{NPMI}}$, and topic diversity) were computed. Second, sampled issues from both major and minor topics were manually inspected to assess topic coherence and classify their contents. Third, intra-rater agreement was measured by blindly re-labeling 300 sampled issues and computing Cohen's $\kappa$.

To investigate longitudinal changes, monthly topic prevalence was computed following Barua et al. \cite{b15}.
To identify significant trends we applied the Mann-Kendall test \cite{b23} with Sen's slope estimator \cite{b24} to each topic's monthly share series, reporting direction, $p$-value, and slope. Shares rather than raw counts are used throughout because corpus volume grows more than tenfold over the study period.

\section{Results}

\subsection{Dataset Overview}

\begin{figure}[htbp]
\centerline{\includegraphics[width=\columnwidth]{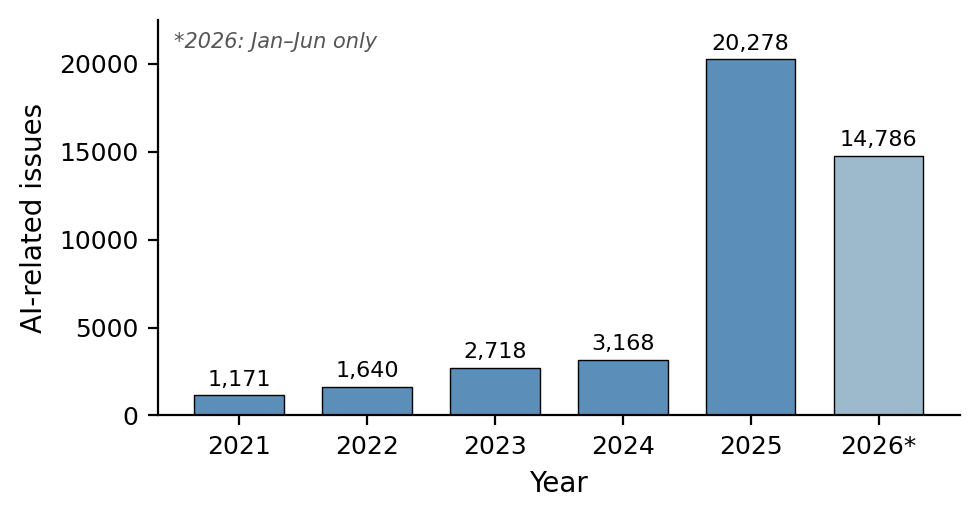}}
\caption{Keyword-retrieved candidate issues per year in microsoft/vscode (2021--2026). 2026 covers 1 January--9 June only. Volume grows more than sixfold between 2024 and 2025, coinciding with agent-mode integration.}
\label{fig:peryear}
\end{figure}

Figure~\ref{fig:peryear} shows the annual number of AI-related issues in the VS Code repository.
The corpus is heavily skewed toward recent years: 1,171 issues in 2021 and 1,640 in 2022, rising to 2,718 (2023), 3,168 (2024), 20,278 (2025), and 14,786 through early June 2026 alone (a partial month, ending 9 June). This inflection coincides with the integration of Copilot Chat and agent-mode features directly into VS Code, which moved AI-related discussion from a peripheral concern to the dominant subject of the project's issue tracker.

\subsection{Topic Model and Validation}\label{sec:val}

Before answering the research questions, we first evaluate the quality of the topic model to ensure that the identified topics are sufficiently coherent and suitable for subsequent analysis.

BERTopic identified 125 topics on the keyword corpus. We establish topic quality before interpreting the topics. BERTopic attains higher coherence than the tuned LDA baseline on every metric: $C_v = 0.628$ vs.\ $0.575$ and $C_{\mathrm{NPMI}} = 0.205$ vs.\ $0.090$ (a 2.3$\times$ gain on the more discriminative normalized PMI measure). LDA's $C_v$ peaked at $k=5$ but varied only narrowly (0.527--0.576) across $k=5$--30, offering little discriminative guidance; its higher topic diversity (0.88 vs.\ 0.39) merely reflects its far smaller topic count (5 vs.\ 125), where word reuse is mechanically rarer, not better separation. We therefore adopt BERTopic as the primary method.

Single-expert inspection complements the metrics. Of the 25 largest topics (Table~\ref{tab:topics}), 18 (72\%) were judged coherent, and across all 125 topics 51 (43.8\% of documents) are AI-substantive (genuinely about generative AI), 27 editor-incidental (matched through incidental keyword co-mention), and the rest noise or unclear. For single-coder reliability, intra-rater test--retest on the full 300-issue blind re-labeling yielded $\kappa = 0.73$ (raw agreement 0.81), indicating substantial self-consistency. We report the keyword-heuristic-versus-content agreement ($\kappa = 0.28$ topic-level, $0.18$ issue-level) only as a measure of keyword--content divergence, not topic validity: its low value shows that keyword surface forms systematically mislabel these discussions (e.g., a topic whose keywords suggest screen recording consists of generic Copilot failure reports), which is precisely why manual content inspection is necessary \cite{b9}, \cite{b17} rather than a formality.

The breadth sample ($n=625$) further shows the long tail is not data garbage: it surfaces fine-grained AI concerns that the largest topics obscure---model-provider fragmentation (distinct topics for GPT-4o/codex, Claude Opus/Sonnet, Gemini, and Grok/DeepSeek), Responsible-AI content filtering and rephrase prompts, agent auto-approval and sandboxing, and billing or refund disputes---indicating that the coarse top-25 view understates the diversity of generative-AI discourse.

\begin{table}[htbp]
\caption{The 25 Largest BERTopic Topics, Grouped Into AI-substantive (A), Editor-incidental (E), and Incoherent (I), With Validated Themes and Coherence (Coh.: Y = coherent). IDs are assigned for readability; the original BERTopic IDs are in the replication package.}
\label{tab:topics}
\begin{center}
\scriptsize
\begin{tabular}{|c|r|l|l|c|}
\hline
\textbf{ID} & \textbf{Size} & \textbf{Validated content} & \textbf{Theme} & \textbf{Coh.} \\
\hline
\multicolumn{5}{|l|}{\textbf{AI-substantive}} \\
\hline
A1 & 1648 & agent file read/edit, sub-agents & code gen. & Y \\
A2 & 966  & agent/chat session mgmt.         & productivity & Y \\
A3 & 664  & coding-agent execution failures  & bug fixing & Y \\
A4 & 641  & MCP server configuration         & productivity & Y \\
A5 & 639  & signature/installation errors    & bug fixing & Y \\
A6 & 615  & file context for AI edits        & code gen. & Y \\
A7 & 608  & chat responsiveness failures     & bug fixing & Y \\
A8 & 590  & quota, billing, subscription     & productivity & Y \\
A9 & 492  & authentication, sign-in          & security & Y \\
\hline
\multicolumn{5}{|l|}{\textbf{Editor-incidental}} \\
\hline
E1 & 2342 & terminal, shell integration      & other & Y \\
E2 & 922  & WSL remote environments          & other & Y \\
E3 & 715  & git, source control              & other & Y \\
E4 & 708  & cursor, pointer, editor UI       & other & Y \\
E5 & 562  & Jupyter notebooks                & other & Y \\
E6 & 508  & chat panel UI/layout             & other & Y \\
E7 & 503  & file explorer UI                 & other & Y \\
E8 & 498  & context menu, window mgmt.       & other & Y \\
E9 & 490  & remote SSH                       & other & Y \\
\hline
\multicolumn{5}{|l|}{\textbf{Incoherent}} \\
\hline
I1 & 732  & triage-bot templates (mixed)     & other & N \\
I2 & 690  & generic Copilot failures         & bug fixing & N \\
I3 & 683  & Copilot account/failures (mixed) & bug fixing & N \\
I4 & 634  & triage-bot templates (mixed)     & other & N \\
I5 & 620  & vague failure reports            & bug fixing & N \\
I6 & 598  & incoherent noise                 & other & N \\
I7 & 589  & generic mixed failures           & bug fixing & N \\
\hline
\end{tabular}
\end{center}
\end{table}

\subsection{RQ1: Topics Discussed}
The primary topic model was deliberately fit on the keyword corpus rather than the filtered one, so that classifier errors could not silently remove themes before manual inspection; the filter enters downstream, in Table~\ref{tab:themes} and in the RQ3 robustness re-clustering. The identified topics reveal that developer discussions are dominated by operational aspects of generative AI rather than conceptual questions.

The topics can be broadly grouped into two categories with incoherent topics (Table~\ref{tab:topics}). The first, and the focus of this study, consists of genuinely AI-related topics, including agent file editing and sub-agent selection (A1), agent and chat session management (A2), coding-agent execution failures (A3), MCP server configuration (A4), file-context handling (A6), chat responsiveness (A7), installation and verification errors (A5), quota and billing (A8), and authentication for Copilot services (A9). These topics show that developers primarily discuss configuring, operating, and troubleshooting AI assistants.

The second category consists of ordinary editor topics, such as terminal, WSL, SSH, Jupyter, source control, cursor behavior, and file explorer issues, which entered the corpus because AI-related keywords appeared incidentally in comments. These topics were therefore excluded from our interpretation of AI-related discussions.

Overall, the results indicate that discussions in the VS Code community focus primarily on the operation and maintenance of AI-assisted development environments rather than on the underlying AI technologies themselves.

\subsection{RQ2: Developer Concerns and Expectations}

Table~\ref{tab:themes} summarizes the theme distribution of the validated sample after semantic filtering. The discussions are overwhelmingly operational in nature. Bug fixing dominates (49.1\%), followed by productivity and workflow concerns (20.7\%) and security (8.9\%).

In contrast, risks frequently emphasized in previous studies remain relatively uncommon in GitHub Issues: hallucination accounts for only 2.4\%, licensing 0.6\%, code generation 1.2\%, and code review is absent. Most reported challenges concern the operation of AI assistants, including execution failures, chat responsiveness, authentication, quota management, and billing, rather than the quality of AI-generated code. Likewise, the observed benefits are primarily reflected in feature requests that improve AI-assisted development workflows.

These findings suggest that developers primarily discuss how to configure, operate, and troubleshoot AI assistants, whereas conceptual risks highlighted in previous literature receive comparatively little attention in the VS Code community. This observation is consistent with previous studies reporting that GitHub Issues mainly capture operational and workflow-oriented discussions \cite{b20}, \cite{b8}.

\begin{table}[htbp]
\caption{Theme distribution of the manually validated sample before and after semantic filtering.}
\label{tab:themes}
\begin{center}
\footnotesize
\begin{tabular}{|l|c|c|}
\hline
\textbf{Theme} & \textbf{Keyword corpus} & \textbf{Filtered corpus} \\
 & \textbf{($n=300$)} & \textbf{($n=169$)} \\
\hline
Bug fixing        & 31.7\% & 49.1\% \\
Productivity      & 12.0\% & 20.7\% \\
Security          & 5.3\%  & 8.9\%  \\
Testing           & 2.0\%  & 3.0\%  \\
Hallucination     & 1.3\%  & 2.4\%  \\
Code generation   & 2.0\%  & 1.2\%  \\
Documentation     & 0.7\%  & 1.2\%  \\
Licensing         & 0.3\%  & 0.6\%  \\
Code review       & 0\%    & 0\%    \\
\hline
Other (incidental) & 44.7\% & 13.0\% \\
\hline
\end{tabular}
\end{center}
\end{table}

\subsection{RQ3: Evolution 2021--2026}\label{sec:rq3}

To investigate how developer discussions evolved over time, we analyzed monthly topic prevalence using the Mann--Kendall trend test. The results consistently show that discussions shifted from traditional editor support toward operational aspects of AI-assisted software development.

Of the 125 topics, 122 contained sufficient data for trend analysis.
On the keyword corpus, 84 of the 122 increase significantly and 17 decrease ($p<0.05$; Tables~\ref{tab:trendsinc} and~\ref{tab:trenddec} report the steepest among the validated coherent topics). As a robustness check, BERTopic was re-fit on the filtered corpus, yielding 86 topics with higher coherence than the primary model ($C_v = 0.648$, $C_{\mathrm{NPMI}} = 0.220$); of these, 78 increase significantly and none decreases. This establishes the rise of AI-operation discussion as the robust longitudinal result and indicates that the apparent declines are relative-share composition artifacts.

As shown in Table~\ref{tab:trendsinc},
the increasing topics are mainly associated with AI-assisted development, including agent and chat session management, file-context handling, installation and verification, and inline chat. These topics emerged after 2023 and became increasingly prominent following the introduction of Copilot Chat and agent-mode features.

As shown in Table~\ref{tab:trenddec},
the decreasing topics mainly correspond to traditional editor functions such as cursor behavior, terminal usage, Jupyter notebooks, and keybindings. Their decline reflects a reduction in relative share as AI-related discussions rapidly expanded rather than an absolute decrease in developer interest. Inline code completion also declines in relative share on the keyword corpus, and unlike the editor topics it is AI-substantive. We did not run a targeted survival test for this topic: the re-clustering produces an independent inventory, and the filtered model's per-document assignments were not retained, so no counterpart can be identified post hoc without re-fitting. What we can state is that no topic decreases significantly in the re-clustered filtered corpus. We therefore treat this decline as a composition effect as well, and rest the shift toward conversational and agent-based interaction on the increasing topics rather than on this decline.

\begin{figure*}[htbp]
\centerline{\includegraphics[width=1.0\textwidth]{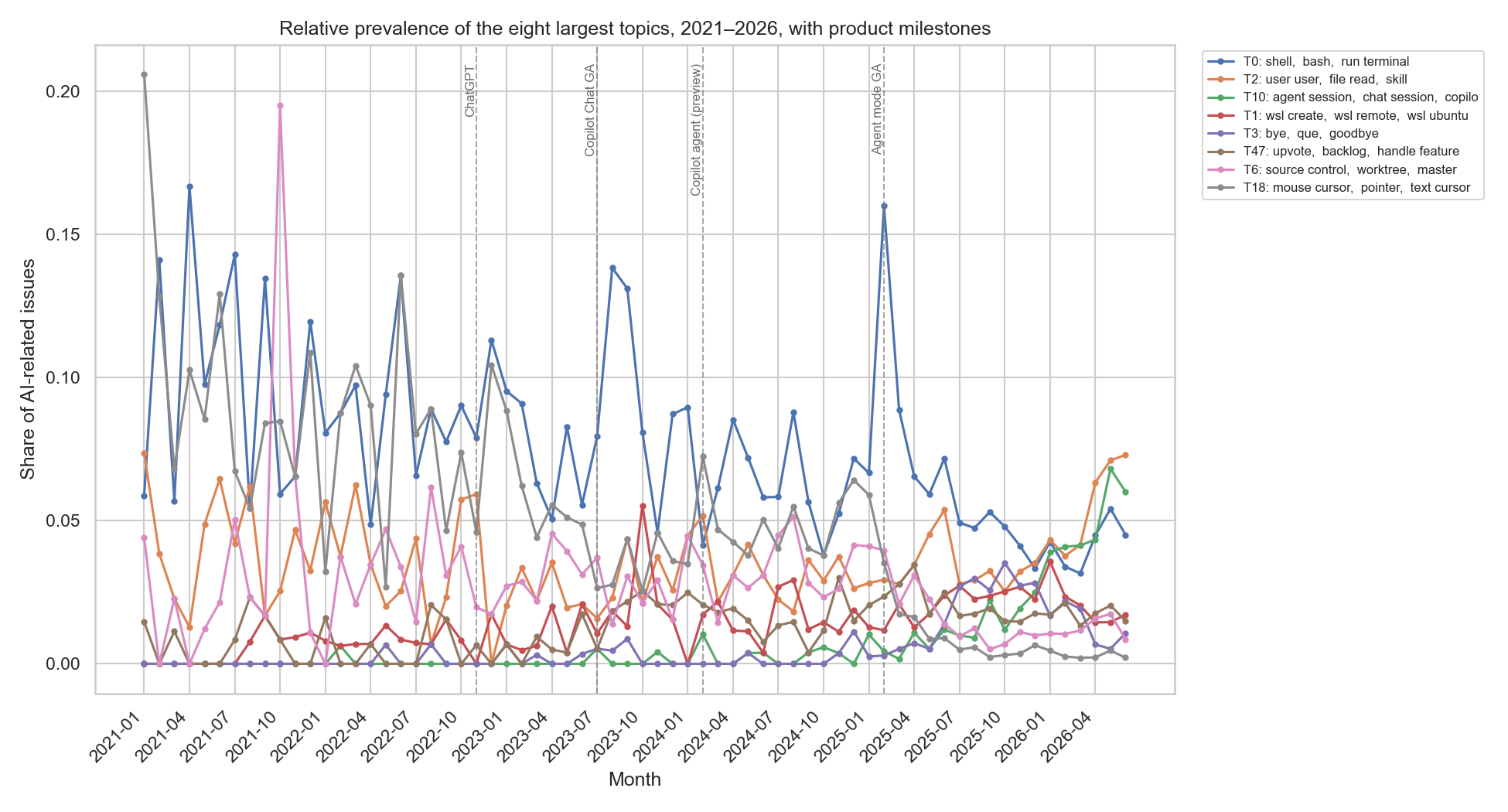}}
\caption{Monthly relative prevalence of the eight largest BERTopic topics, 2021--2026. Dashed lines mark product milestones: ChatGPT (Nov 2022), Copilot Chat GA (2023), Copilot agent preview (2024), and agent mode GA (2025).}
\label{fig:prev}
\end{figure*}

Figure~\ref{fig:prev} illustrates this transition. AI-related discussions evolved through three phases: early discussions centered on Copilot as a code-completion tool (2021--2022), expanded with Copilot Chat (2023--2024), and became dominated by agent-based development and ecosystem integration from 2025 onward.

Overall, the results indicate a clear evolution in developer attention, from using AI primarily for code completion to adopting AI assistants as configurable and increasingly autonomous development partners.

\begin{table}[htbp]
\caption{Mann-Kendall Results: Steepest Significant Increasing Trends Among Validated Coherent Topics (Monthly Share, $p<0.05$).}
\label{tab:trendsinc}
\begin{center}
\scriptsize
\begin{tabular}{|c|l|c|r|}
\hline
\textbf{\#} & \textbf{Topic} & \textbf{$p$} & \textbf{Slope ($\times10^{-3}$)} \\
\hline
1 & WSL remote (Copilot in WSL)        & $<$0.001 & $+0.31$ \\
2 & installation / verification errors & $<$0.001 & $+0.29$ \\
3 & file context for AI edits          & $<$0.001 & $+0.24$ \\
4 & inline chat panel / layout         & $<$0.001 & $+0.23$ \\
5 & agent / chat session management    & $<$0.001 & $+0.17$ \\
\hline
\end{tabular}
\end{center}
\end{table}

\begin{table}[htbp]
\caption{Mann-Kendall Results: Steepest Significant Decreasing Trends Among Validated Coherent Topics (Monthly Share, $p<0.05$). These are relative-share composition effects; no decreasing trend reappears in the robustness re-clustering.}
\label{tab:trenddec}
\begin{center}
\scriptsize
\begin{tabular}{|c|l|c|r|}
\hline
\textbf{\#} & \textbf{Topic} & \textbf{$p$} & \textbf{Slope ($\times10^{-3}$)} \\
\hline
1 & mouse cursor, pointer             & $<$0.001 & $-1.56$ \\
2 & inline suggestion / autocomplete  & $<$0.001 & $-0.97$ \\
3 & terminal, shell integration       & $<$0.001 & $-0.94$ \\
4 & Jupyter notebooks                 & $<$0.001 & $-0.76$ \\
5 & keybindings                       & $<$0.001 & $-0.72$ \\
\hline
\end{tabular}
\end{center}
\end{table}

\section{Discussion}

Our findings suggest that discussions in GitHub Issues differ substantially from the concerns emphasized in much of the existing literature. While previous studies often focus on hallucination, licensing, and code-quality risks \cite{b3}, \cite{b4}, \cite{b21}, \cite{b22}, the VS Code community primarily discusses practical issues related to deploying and operating AI assistants, such as configuration, authentication, billing, and reliability. This suggests that surveys and issue trackers capture complementary aspects of developer experience and should not be treated as interchangeable data sources.

Another notable finding is the changing role of generative AI within the development environment. Over the study period, discussions shifted from code-completion features toward AI assistants integrated into everyday development workflows. Developers increasingly treat AI assistants as part of the development infrastructure rather than as standalone productivity tools.

\section{Threats to Validity}\label{sec:threats}
\textit{Construct validity.} Keyword retrieval is imprecise---44.7\% of sampled issues co-mention an AI term incidentally---because the dominant noise is semantic (editor bugs that genuinely contain ``copilot''), which a lexical proximity filter cannot remove (precision ceiling 0.60). The semantic relevance classifier (Section~\ref{sec:collect}) is our primary mitigation, cutting incidental issues to 13.0\% of the labeled sample at precision 0.79; its 0.84 recall means some relevant issues are still dropped, and excluding ``GPT'' from server-side search (retained only as a post-filter) may further cost recall. The Search API's 1,000-result cap is mitigated by monthly windows with recursive splitting.

\textit{Internal validity.} This is a single-annotator study, so topic validity rests primarily on annotator-independent coherence metrics, supported by single-expert coherence judgments and intra-rater test--retest reliability rather than inter-coder agreement. The content labels were AI-assisted; since the topics come from a neural model and the labels from a transformer-based assistant, the two could share representational priors, so every label was reviewed and, where needed, overridden by the first author, and the blind re-labeling pass used for intra-rater $\kappa$ provides a check that the AI assistance did not simply propagate the model's groupings. The residual threat---the subjectivity of a single coder---is not fully removable without a second independent human annotator, which we identify as the priority next step.

\textit{External validity.} VS Code is a single Microsoft-owned repository that ships Copilot as a first-party feature; its profile may not generalize to projects without first-party AI integrations.

\textit{Conclusion validity.} Mann-Kendall flags emergence (zero-then-nonzero series) as monotonic increase, so we distinguish emergent from trending topics, and we treat the decreasing trends---which do not survive the robustness re-clustering---as composition artifacts rather than findings.

\section{Conclusion}

This paper presented a longitudinal analysis of generative-AI discussions in the VS Code GitHub repository, covering 43,806 candidate issues retrieved from 2021 to 2026, of which 25,227 were retained by semantic filtering for theme validation and a robustness re-clustering. The results show that developer discussions are dominated by operational concerns, such as configuration, reliability, authentication, and workflow integration, while conceptual risks frequently discussed in previous studies receive comparatively little attention. Over time, discussions shifted from AI-assisted code completion toward conversational and agent-based development, reflecting the increasing integration of generative AI into everyday software development.

These findings provide empirical evidence that GitHub Issues capture a practical, workflow-oriented perspective on generative AI that complements survey-based studies and other developer communities.

Future work includes validating the findings across additional OSS projects and extending the analysis using complementary techniques such as sentiment analysis and alternative topic-modeling methods.

%\section*{Acknowledgment}
%The authors thank Prof.\ Akito Monden for supervision and guidance.

\end{document}